\documentclass[conference]{IEEEtran}
\usepackage[dvipdf]{graphicx}
\usepackage{amsmath}
\usepackage{subcaption}
\usepackage{hyperref}
\usepackage{textcomp}

\graphicspath{{./tikzImages/}}

\begin{document}

\title{Delay Line as a Chemical Reaction Network}

\author{\IEEEauthorblockN{Josh Moles}
\IEEEauthorblockA{Department of Electrical and\\
Computer Engineering\\
Portland State University\\
jmoles@pdx.edu}
\and
\IEEEauthorblockN{Peter Banda}
\IEEEauthorblockA{Department of Computer Science\\
Portland State University\\
banda@pdx.edu}
\and
\IEEEauthorblockN{Christof Teuscher}
\IEEEauthorblockA{Department of Electrical and\\
 Computer Engineering\\
Portland State University\\
teuscher@pdx.edu}}

\maketitle

\begin{abstract}
Chemistry as an unconventional computing medium presently lacks a systematic approach to gather, store, and sort data over time. To build more complicated systems in chemistries, the ability to look at data in the past would be a valuable tool to perform complex calculations. In this paper we present the first implementation of a chemical delay line providing information storage in a chemistry that can reliably capture information over an extended period of time. The delay line is capable of parallel operations in a single instruction, multiple data (SIMD) fashion.

Using Michaelis-Menten kinetics, we describe the chemical delay line implementation featuring an enzyme acting as a means to reduce copy errors. We also discuss how information is randomly accessible from any element on the delay line. Our work shows how the chemical delay line retains and provides a value from a previous cycle. The system's modularity allows for integration with existing chemical systems. We exemplify the delay line capabilities by integration with a threshold asymmetric signal perceptron to demonstrate how it learns all 14 linearly separable binary functions over a size two sliding window. The delay line has applications in biomedical diagnosis and treatment, such as smart drug delivery.
\end{abstract}


\section{Introduction}
The ability to store temporal data is a fundamental operation to many types of calculations and signal processing~\cite{kanopoulos1986first}. Access to the results from previous calculations or observations is essential to form more complex operations and devices, like first-in, first-out (FIFO) memories. Capturing the values of data over time in a chemical reaction network could enable sensing of concentrations over a window of time rather than just the present concentration observed. Chemistry provides an unconventional paradigm to solve computation tasks in a highly parallel fashion because the reactions changing the concentration of species inherently occur concurrently~\cite{de2003multitasking}. 

Chemical computing is rapidly growing field that would benefit from such a time delay line. Previous works on chemical computers address problems such as networking protocols~\cite{Meyer2011}, logic circuits~\cite{Matsumaru2005}~\cite{Banda2013}~\cite{Arkin1994}, signal processing~\cite{Jiang2013}, tic-tac-toe~\cite{stojanovic2003deoxyribozyme} and chess~\cite{faulhammer2000molecular}. Arkin and Ross also discuss the need for a ``buffer'' between the phases of logic elements described in their paper~\cite{Arkin1994}. All of these areas could benefit from a time delay line implemented in a chemistry. Most of these systems only have the capability to look into the present or previous value as inputs. With our delay line, the period of looking into the past could be much longer.

An example biochemical application is smart medication~\cite{neat1988hybrid}~\cite{abbod2002survey}. Rather than have a fixed dosage of a specific type of medicine, a patient could be observed over a time window and then adapt the drug (in quantity or species) to best respond to their needs. Another use in the biochemistry field would be the detection of harmful species, e.g., chemicals produced by cancer cells in a host. With a time delay line, the detection would not be limited to a simple yes or no, but can get extended to measure a chemical concentration as well as capture at what point the event occurred. This could be used for monitoring or passed to another chemical computer (such as a chemical perceptron) that could then make decisions and react.

This work presents two variations of a time delay line implemented as a chemical reaction network. First, a manual copy model that requires additional signaling to indicate when it is time to propagate values through the delay line and the other featuring an automatic backwards propagation of a copy signal. We will also demonstrate an example of this delay line by connecting it to a chemical perceptron capable of learning the 14 lineally separable two-input logic functions~\cite{Banda2013}. 

The buffer presented in this work is novel compared to prior work. Jiang et al.~\cite{Jiang2013} introduced the concept of a delay element. The delay element is primarily used as a storage area for holding data in between each computation cycle. The data then comes back and is examined in computing during the next iteration of the calculation. Jiang's buffer is primarily a signal processing application looking only at the previous value. Our delay line has the ability to delay not only multiple steps in time, but also allows access to any of the past values besides the most recent. An example implementation could remove the intermediate stages of the delay line and only look at the final output. This would create a FIFO~\cite{kanopoulos1986first}.

Other areas, such as networking, use chemical reaction networks as a mechanism to control scheduling and queuing of packets~\cite{Meyer2011}. The work discusses a methodology to use the law of mass action as a means to schedule packets. With a buffer like the one we are describing, the work could also be extended to actually implement a means to queue packets in a chemical system. This system would reduce cost and complexity by having a single implementation medium.

This paper will start by providing an overview of chemical reaction networks to understand the concepts discussed next in the design (Section~\ref{sec:delayline}) and results (Section~\ref{sec:results}) sections. We then describe a deoxyribozyme cascading implementation (Section~\ref{sec:deoxy}) followed by our concluding remarks (Section~\ref{sec:conclude}).

\section{Artificial Chemistry Background}
The chemical system presented here uses several observations of nature (such as chemical kinetics and the law of conservation of mass) to model the behavior of the delay line. We follow the standard formalism for representing chemistry called chemical reaction network, which is an instance of an artificial chemistry~\cite{Dittrich2001}. It consists of a set of species and reactions with associated rates. In our system, we assume molecular species are symbolic and unstructured. There is no notion of space because we assume the solution is well stirred. We do not need to handle the position of an individual module, but rather transform all molecules of the same type (species) using rates generated by kinetic laws: mass-action~\cite{horn1972general}~\cite{erdi1989mathematical} for regular and Michaelis-Menten~\cite{henri1903lois}~\cite{Michaelis1913}~\cite{Leskovac2003} for catalytic reactions.

Dittrich~\cite{Dittrich2001} describes an artificial chemistry made up of a finite set of molecular species and a finite set of reactions. The set of molecular species are represented by symbols (e.g., $X$, $X_1$, $Y_1$, $X1_{signal}$). The reactions are formed through multiple sets of species (reaction left side) that react to form products (reaction right side)~\cite{Banda2013}. A reaction looks like $X_1 + X_2 \rightarrow Y$ where reactants $X_1$ and $X_2$ form the product $Y$. 

We combine mass action kinetics with the ideas of artificial chemistry to express reaction rates for ordinary (non-catalytic) reactions. Epstein~\cite{Epstein1998} expresses this through a series of differential equations. Given a generic chemical reaction $aX_1 + bX_2 \rightarrow cY$, the rate of reaction, $v$, is expressed by
\begin{equation}
v = -\frac{1}{a}\frac{d[X_1]}{dt} = -\frac{1}{b}\frac{d[X_2]}{dt} = \frac{1}{c}\frac{d[Y]}{dt} = k[X_1]^a[X_2]^b,
\end{equation}
where $[X_1]$, $[X_2]$, and $[Y]$ are the concentrations of the reactants, $X1$ and $X2$, and the product, $Y$. Symbols $a$ and $b$ are stoichiometric constants, and $k$ is the reaction rate constant. Reactions could also be reversible, but in this paper, for simplification, we assume the reverse rate is always zero.

Michaelis-Menten kinetics describes the rate of a catalytic reaction where a substrate ($S$) is transformed into a product ($P$) through the use of an enzyme or catalyst($E$) in a reaction modeled as $E + S \rightleftharpoons ES \rightarrow E + P$. The catalyst $E$ speeds up the rate of the reaction without being consumed in the process. The reaction rate for this type of reaction is defined by
\begin{equation}
v = \frac{k_{cat}[E][S]}{K_m+[S]},
\end{equation}
where $k_{cat}$ and $K_m$ are rate constants. Copeland~\cite{copeland2004enzymes} discusses the work of Henri, Michaelis, and Menten in greater detail.

\section{Delay Line Design}
\label{sec:delayline}
To introduce the time delay line design, we will first examine a delay line constructed of only two stages in two different styles. One is a manual copy delay line that requires experimenter participation to indicate when it is time to move values between stages. The second model automatically propagates the signaling species backwards, hence it is more autonomous, but it comes at the cost of additional and cumulative error in the resulting output values.

\subsection{Manual Copy Delay Line}
\begin{figure}[h]
	\centering
	\includegraphics[width=5cm]{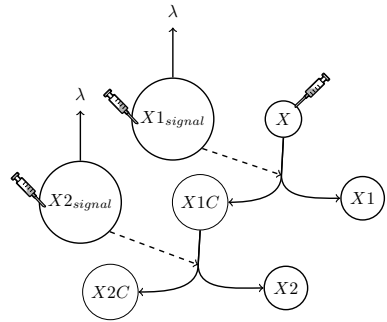}
	\caption{Manual copy delay line with two stages. The syringe is used to indicate the species where inputs are presented and $X1$ and $X2$ represent the output species from the delay line. Species $X2C$ is used to cascade a value to a delay line of greater than two stages. $X1_{signal}$ and $X2_{signal}$ catalyze the copy and decay ($\lambda$).}
	\label{fig:manualprop_n_2}
\end{figure}

First, we will discuss the delay line of two stages with manual copy of the signaling species. This element is shown in Figure~\ref{fig:manualprop_n_2}. A delay line of two stages is composed of seven species: $X$, $X1C$, $X1$, $X2$, $X2C$, $X2_{signal}$, and $X1_{signal}$. The species $X$ represents the input of value to the delay line. The signals, $X1_{signal}$ and $X2_{signal}$, are the catalysts that start the reaction conversion of $X$ into corresponding stages. The primary function of $X1_{signal}$ is to trigger and accelerate the copy reaction which does the conversion of $X$ to $X1C$ and $X1$. Species $X2_{signal}$ performs a similar action for the conversion of $X1C$ to $X2$.

Species $X1C$ and $X2C$ are delayed copies of $X$ that are used to copy to the next stage of the system (for example, $X1$ to $X2$ and $X2C$). Species $X2C$ is shown for completeness and is used to cascade the system to a delay line of more than two stages. For a two stage delay line, it is waste and flushed. The outputs of the system are the $X1$ and $X2$ species. $X1$ and $X2$ represent the current and previous values of $X$ that are consumed as the inputs of another system.

Internal to the system is the transition storage species, $X1C$. The storage species acts as a buffer for the value that will transition into $X2$ on next activation of the system with an $X2_{signal}$ passed in. Ideally, the concentration of $X1C$ will be the same as $X1$ prior to its consumption. This process is represented by a set of reactions using the previously mentioned species. Reactions~\ref{eq:XreacMan}~and~\ref{eq:X1CreacMan} represent the conversion of the input species, $X$, through to the output species, $X1$ and $X2$. 
\begin{alignat}{2}
X & \xrightarrow{X1_{signal}} & X1 + X1C  \label{eq:XreacMan} \\
X1C & \xrightarrow{X2_{signal}} & X2 + X2C \label{eq:X1CreacMan}
\end{alignat}
Reactions~\ref{eq:X2sigMan} and \ref{eq:X1sigMan} show the decay (represented by lambda, $\lambda$) of the catalyst species, $X1_{signal}$ and $X2_{signal}$.
\begin{alignat}{2}
X2_{signal} & \rightarrow & \lambda \label{eq:X2sigMan} \\ 
X1_{signal} & \rightarrow & \lambda \label{eq:X1sigMan}
\end{alignat}

Now, using these reactions, we can examine data moving through the delay line. For this manual copy delay line, actions must occur at two moments (in time). First, at time zero, we present a random value to the input $X$ and reset $X1$ and $X2$ to zero. These are set to zero to simulate consumption by the underlying system the delay line is integrated with. Species $X2_{signal}$ is set to one to copy the value stored in $X1C$ to $X2$. In the ideal case for the initialization and first run of the delay line, $X2$ should be zero until these actions repeat. After 25 time steps, $X1_{signal}$ is injected to the system. The wait is to fully allow the transition of $X1C$ to $X2$ before beginning the reaction of $X$ to $X1C$. These injections repeat every 1,000 time steps and are summarized in Table~\ref{tab:as2Man}. Table~\ref{tab:as2Pipeline} shows an example of these these injections repeating every 1,000 time steps with example data moving through.

\begin{table}[h]
	\caption{Actions for two stage manual copy delay line simulations.}
	\label{tab:as2Man}
	\centering
	\begin{tabular}{lll}
		Time		& Species			& Value 					\\ \hline
		0			& $X$				& $0.0 \le rand() \le 1.0$	\\
		0			& $X1$				& 0							\\
		0			& $X2$				& 0							\\
		0			& $X2_{Signal}$ 	& 1							\\
		25			& $X1_{Signal}$ 	& 1							\\
	\end{tabular}
\end{table}

\begin{table}
	\centering
	\caption{Pipeline view of data moving through chemistry from Table~\ref{tab:as2Man}. Bold items show those injected to the system. \textbf{A}, \textbf{B}, and \textbf{C} are inputs and \textbf{1} is a concentration (presence) of $Xm_{signal}$.}
	\label{tab:as2Pipeline}	
    \begin{tabular}{l|l|l|l|l|l|l}
    Species  & Time=0                     & 25                         & 1000                       & 1025                       & 2000                       & 2025                       \\ \hline
    X        & $\textbf{A}$               & $A \rightarrow 0$          & $\textbf{B}$               & $B \rightarrow 0$          & $\textbf{C}$               & $C \rightarrow 0$          \\
    X1signal & ~                          & $\textbf{1} \rightarrow 0$ & ~                          & $\textbf{1} \rightarrow 0$ & ~                          & $\textbf{1} \rightarrow 0$ \\
    X2signal & $\textbf{1} \rightarrow 0$ & ~                          & $\textbf{1} \rightarrow 0$ & ~                          & $\textbf{1} \rightarrow 0$ & ~                          \\ \hline
    X1       & $\textbf{0}$               & $0 \rightarrow A$          & $\textbf{0}$               & $0 \rightarrow B$          & $\textbf{0}$               & $0 \rightarrow C$          \\
    X1C      & ~                          & $\rightarrow A$            & $A \rightarrow 0$          & $0 \rightarrow B$          & $B \rightarrow 0$          & $0 \rightarrow C$          \\
    X2       & $\textbf{0}$               & ~                          & $\textbf{0} \rightarrow A$ & $A$                        & $\textbf{0} \rightarrow B$ & $B$                        \\
    X2C      & ~                          & ~                          & $\rightarrow A$            & $A$                        & $A \rightarrow B$          & $B$                        \\
    \end{tabular}
\end{table}

\begin{figure*}[!t]
	\centering
	\begin{subfigure}[b]{0.45\textwidth}
		\centering
		\includegraphics[width=5cm]{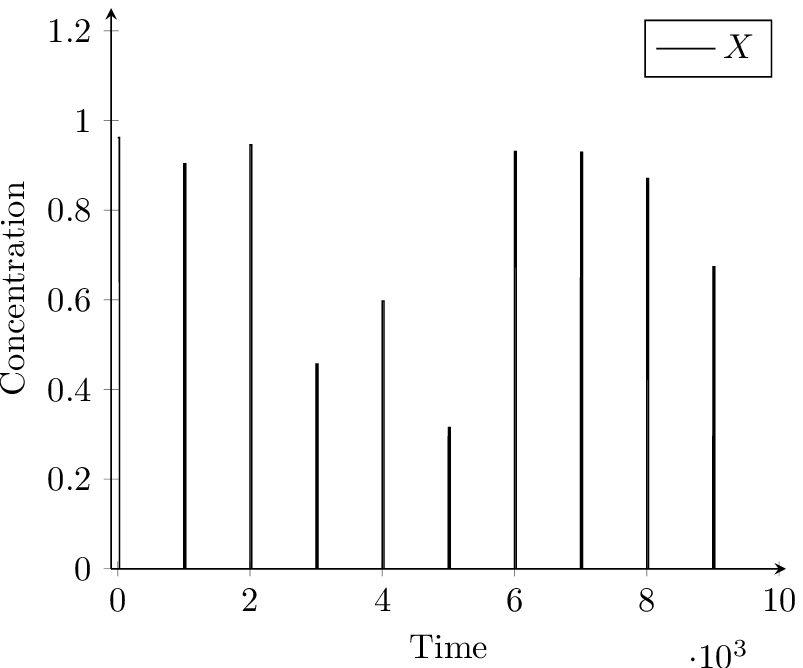}
		\caption{Input}
		\label{fig:delay2XMan}
	\end{subfigure}
	~
	\begin{subfigure}[b]{0.45\textwidth}
		\centering
		\includegraphics[width=5cm]{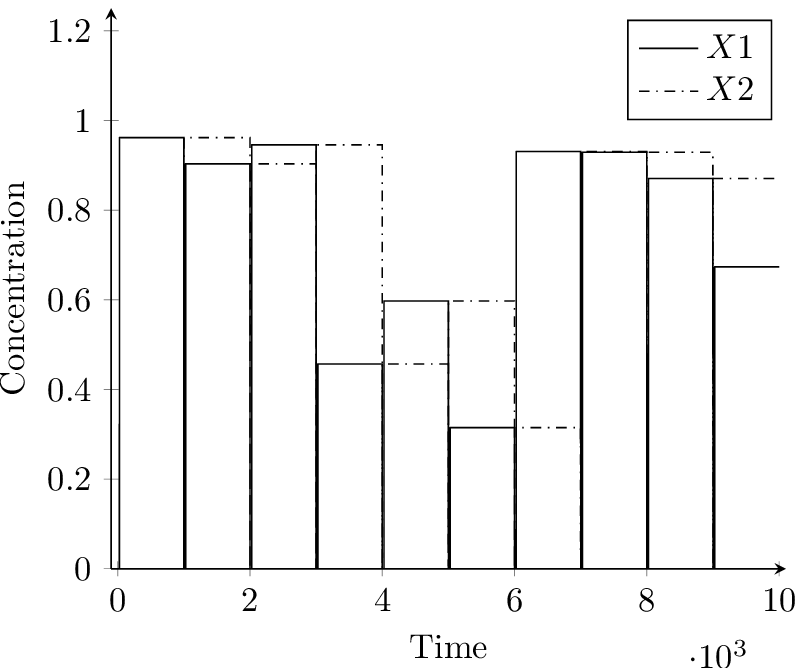}
		\caption{Outputs}
		\label{fig:delay2X1Man}
	\end{subfigure}
	\\
	\begin{subfigure}[b]{0.45\textwidth}
		\centering
		\includegraphics[width=5cm]{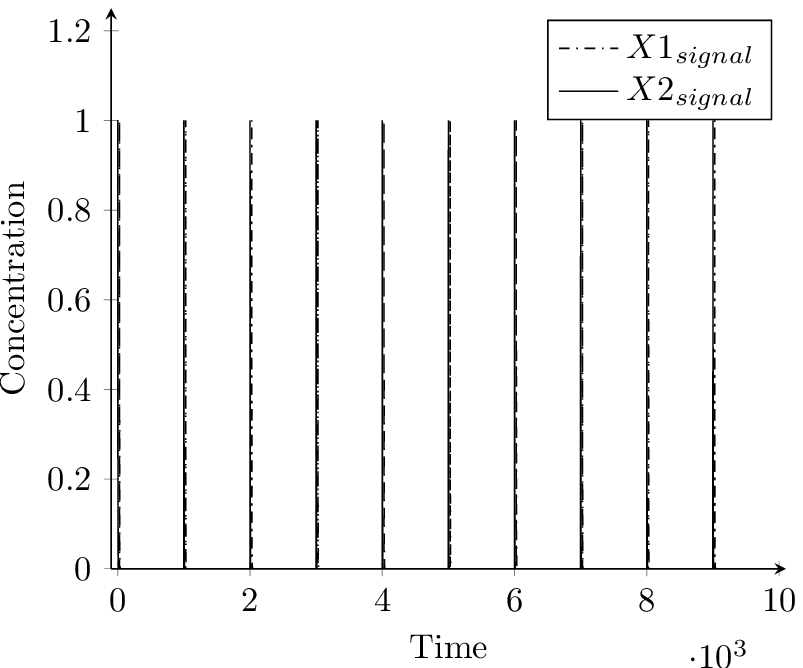}
		\caption{Copy Signals}
		\label{fig:delay2X2SignalMan}
	\end{subfigure}
	~
	\begin{subfigure}[b]{0.45\textwidth}
		\centering
		\includegraphics[width=5cm]{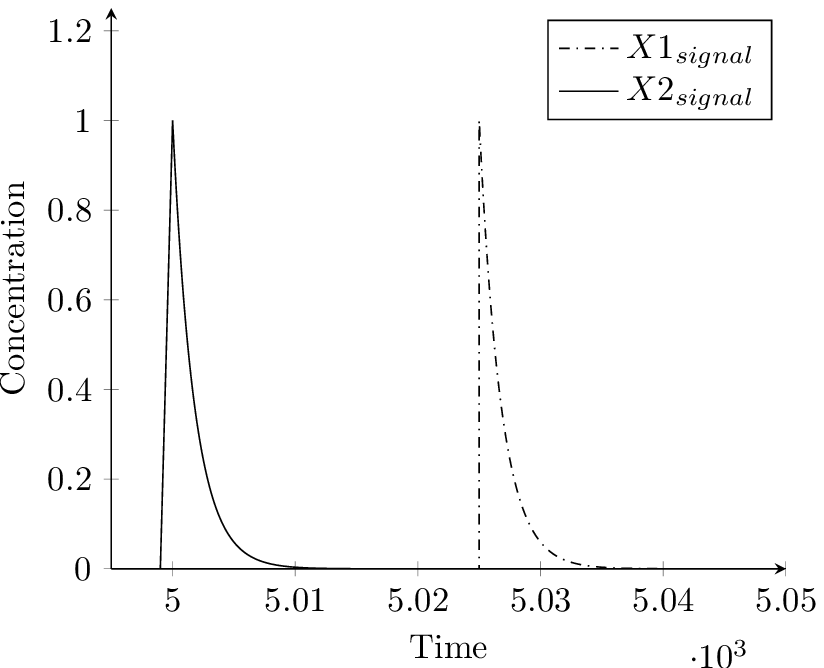}
		\caption{Copy Signals (Zoomed on Figure~\ref{fig:delay2X2SignalMan})}
		\label{fig:delay2X2SignalZoomedMan}
	\end{subfigure}
	\caption{Two stage manual copy delay line showing inputs and outputs. Data arrives as input (\ref{fig:delay2XMan}) and is available on outputs (\ref{fig:delay2X1Man}) with $X1$ being the current and $X2$ being the previous $X$. The copy of this data is triggered by $X1_{signal}$ and $X2_{signal}$ (\ref{fig:delay2X2SignalMan}). Figure~\ref{fig:delay2X2SignalZoomedMan} shows the copy signals zoomed in from Figure~\ref{fig:delay2X2SignalMan}.}
	\label{fig:delay2allMan}
\end{figure*}

Figure~\ref{fig:delay2allMan} shows the results of running the actions in Table~\ref{tab:as2Man} for 10 iterations (10,000 time steps). Valid data is available for examination on output species $X1$ and $X2$ every time steps after each cycle. Figure~\ref{fig:delay2XMan} shows the input values injected to this manual delay line. During the first cycle, species $X2$ remains at zero since there is no previous value as seen in Figure~\ref{fig:delay2X1Man}. Figure~\ref{fig:delay2X2SignalMan} shows the catalysts, $X2_{signal}$ and $X1_{signal}$, sequentially getting injected each cycle. Looking on a zoomed in view in Figure~\ref{fig:delay2X2SignalZoomedMan}, the sequence of actions where $X2_{signal}$ is injected at time zero followed by $X1_{signal}$ 25 time steps later.

\subsection{Backwards Signal Propagation Delay Line}
\begin{figure}[!t]
	\centering
	\includegraphics[width=5cm]{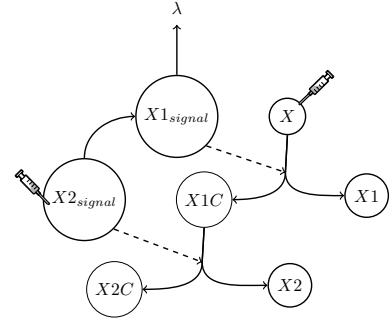}
	\caption{Backwards propagating delay design with two stages. The syringe is used to indicate an injection of the input species $X$ and the copy signal $X2_{signal}$. The species $X1$ and $X1$ represent the output species from the delay line. The signal $X2_{signal}$ is propagated backwards to $X1_{signal}$ without user intervention and then decays ($\lambda$).}
	\label{fig:delayelement}
\end{figure}

The backwards signal propagation delay line handles the signal species differently. More specifically, the only signaling species is $X2_{signal}$ and rather than decay, $X2_{signal}$ reacts to $X1_{signal}$. The advantage of this model is that the user is only required to perform actions at the beginning of the cycle and then the system transforms the species internally (without external help). A revised figure of this change is shown in Figure~\ref{fig:delayelement}. This reduces the number of injections to two: the input ($X$) and the final copy signal ($X2_{signal}$ for two stage). The change leaves reactions~\ref{eq:Xreac} and \ref{eq:X1Creac} unchanged.

\begin{alignat}{2}
X & \xrightarrow{X1_{signal}} & X1 + X1C  \label{eq:Xreac} \\
X1C & \xrightarrow{X2_{signal}} & X2 + X2C\label{eq:X1Creac}
\end{alignat}

Revising the remaining reactions requires modifying only reaction~\ref{eq:X2sigMan}. Removing the decay from reaction~\ref{eq:X2sigMan} so that $X2_{signal}$ reacts to $X1_{signal}$ gives the updated reactions~\ref{eq:X2sig} and \ref{eq:X1sig}.

\begin{alignat}{2}
X2_{signal} & \rightarrow & X1_{signal} \label{eq:X2sig} \\ 
X1_{signal} & \rightarrow & \lambda \label{eq:X1sig}
\end{alignat}

All actions in the system occur instantaneously and are the same as actions employed by the manual delay line at time zero. At the beginning of every cycle, $X1$ and $X2$ are set to zero to simulate the next block of the system consuming their values. A random value is presented on the input of the system, $X$, and $X2_{signal}$ is presented to start the system. Table~\ref{tab:as2} summarizes these steps. These actions repeat every 1,000 time steps to ensure enough time for all reactions to reach steady state.

\begin{table}
	\caption{Actions for two stage back propagation delay line simulations. These actions repeat every 1000 similar to Table~\ref{tab:as2Pipeline}.}
	\label{tab:as2}
	\centering
	\begin{tabular}{lll}
		Time		& Species			& Value 					\\ \hline
		0			& $X$				& $0.0 \le rand() \le 1.0$	\\
		0			& $X1$				& 0							\\
		0			& $X2$				& 0							\\
		0			& $X2_{Signal}$ 	& 1							\\
	\end{tabular}
\end{table}

\begin{figure*}[!t]
	\centering
	\begin{subfigure}[b]{0.45\textwidth}
		\centering
		\includegraphics[width=5cm]{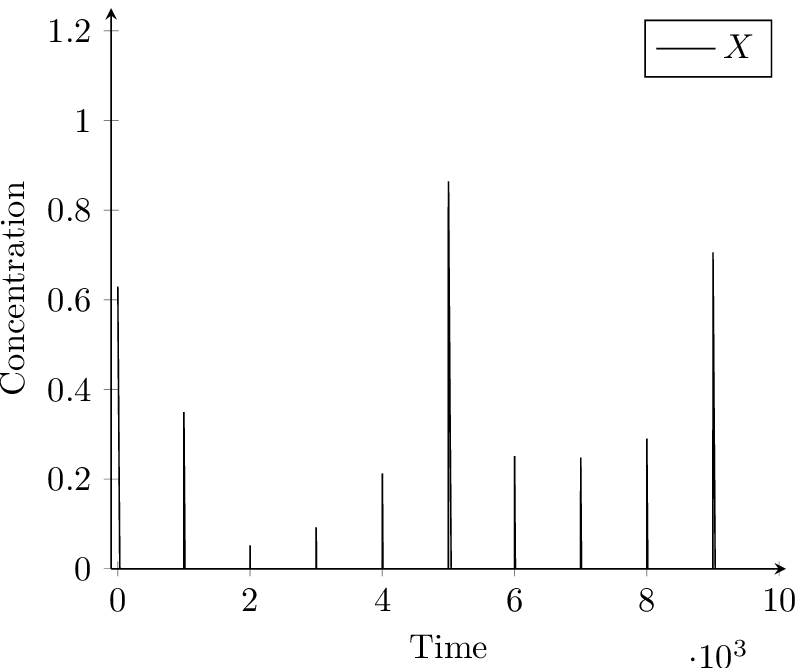}
		\caption{Input}
		\label{fig:delay2X}
	\end{subfigure}
	~
	\begin{subfigure}[b]{0.45\textwidth}
		\centering
		\includegraphics[width=5cm]{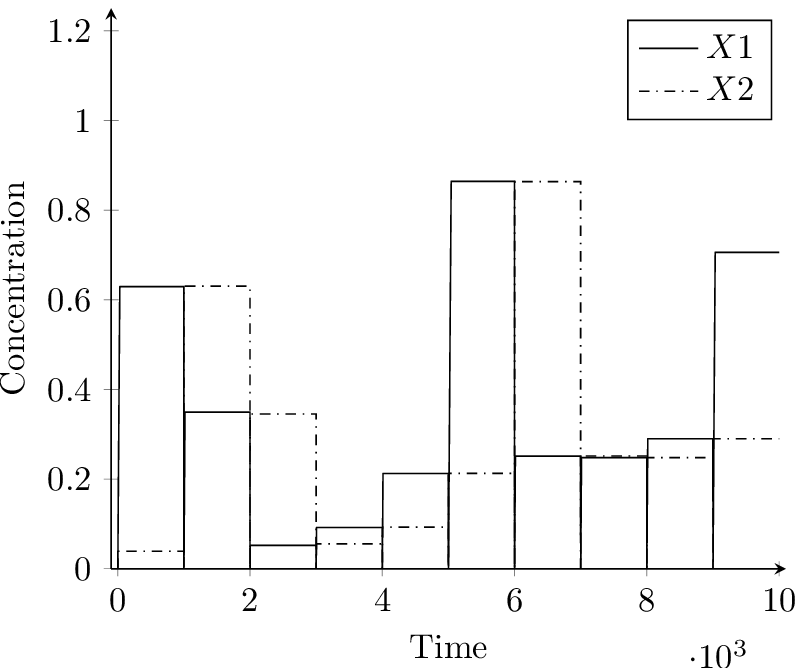}
		\caption{Outputs}
		\label{fig:delay2X1}
	\end{subfigure}
	\\
	\begin{subfigure}[b]{0.45\textwidth}
		\centering
		\includegraphics[width=5cm]{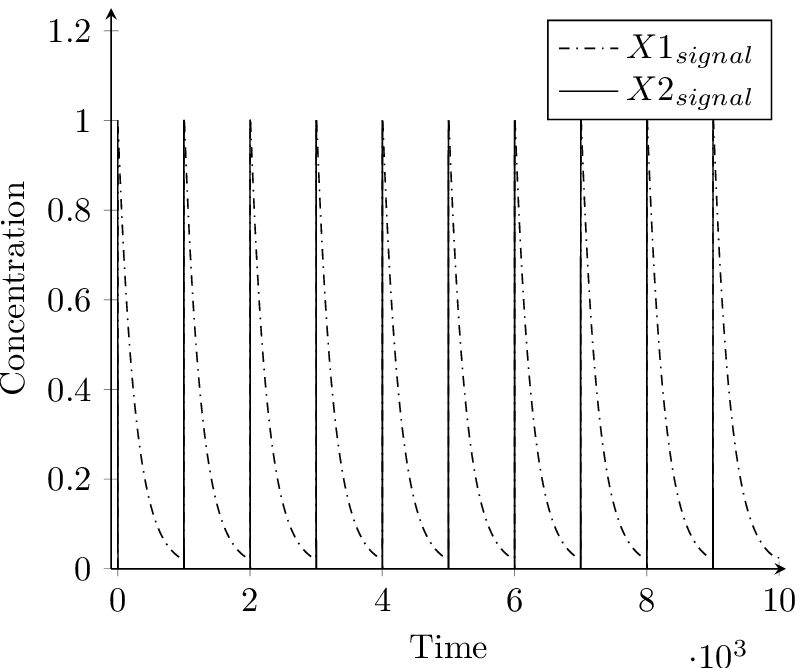}
		\caption{Copy Signals}
		\label{fig:delay2X2Signal}
	\end{subfigure}
	~
	\begin{subfigure}[b]{0.45\textwidth}
		\centering
		\includegraphics[width=5cm]{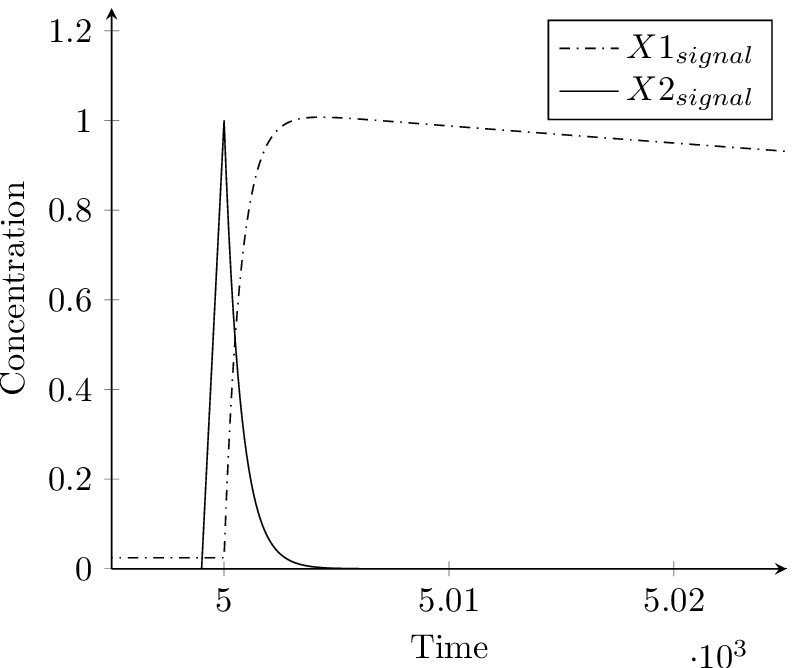}
		\caption{Copy Signals (Zoomed on Figure~\ref{fig:delay2X2Signal})}
		\label{fig:delay2X2SignalZoomed}
	\end{subfigure}
	\caption{Two stage backwards propagation delay line showing inputs and outputs. Data arrives as input (\ref{fig:delay2X}) and is available on outputs (\ref{fig:delay2X1}) with $X1$ being the current and $X2$ being the previous $X$. The copy is started by $X1_{signal}$ and $X2_{signal}$ (\ref{fig:delay2X2SignalMan}). Figure~\ref{fig:delay2X2SignalZoomedMan} shows the signals controlling propagation zoomed in from Figure~\ref{fig:delay2X2SignalMan}.}
	\label{fig:delay2all}
\end{figure*}

The simulations of the backwards propagation delay line run for 10,000 time steps (same as for the manual delay line). Valid data is also produced at the same point (every 50 steps) on the output species $X1$ and $X2$. The value produced on the first cycle of $X2$ ideally should be zero, but leakage from $X1C$ is generally seen from steps zero to 1,000 (see Figure~\ref{fig:delay2X1}). An input is introduced to the system at species $X$ (Figure~\ref{fig:delay2X}) and then is reacted in the same cycle to species $X1$ (Figure~\ref{fig:delay2X1}). After the next cycle (i.e., the next introduction of $X2_{signal}$), the value injected at $X$ previously is now presented at $X2$ (Figure~\ref{fig:delay2X1}).

Notice that the backwards propagation introduces an error to the system with some of the $X2$ values not lining up exactly with the previous $X1$. This difference is due to the window of time that both reactions for the copy of $X$ are active simultaneously. Looking at Figure~\ref{fig:delay2X2SignalZoomed}, $X1_{signal}$ and $X2_{signal}$ are large enough for both catalyses to occur. So, for this small window of time, there is effectively a direct path from $X$ to cascade down to $X2$. This overlap is not inherently a problem. It allows the desired parallelism of this system. We can afford this error in a small number of stages, but the inaccuracy can grow with a larger number of stages.

\subsection{Inherit Single Instruction, Multiple Data (SIMD)}
\begin{figure}[!t]
	\centering
	\includegraphics[width=5cm]{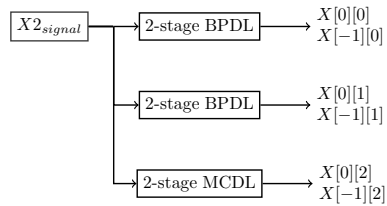}
	\caption{Time delay design single instruction, multiple data (SIMD) representation showing simultaneous output of previous ($X[-1][n]$) and current ($X[0][n]$) $X$ for parallel data processing. The signaling can be used with multiple instances of a delay line, both for the manual copy and the back propagation type.}
	\label{fig:simd}
\end{figure}

With the nature of chemistry, one of the advantages of our unconventional delay line implementation is the ability to perform single instruction, multiple data~\cite{flynn1972some} operations. The main factor is finding a unique set of species to hold each delay line that will not react with surrounding buffers to allow such parallel operations. Figure~\ref{fig:simd} shows an example of a two-stage set of backwards propagation and manual copy delay lines that are producing a vector of three values for the current and previous cycles.

\subsection{More than Two Stages}
\begin{figure}[!t]
	\centering
	\includegraphics[width=5cm]{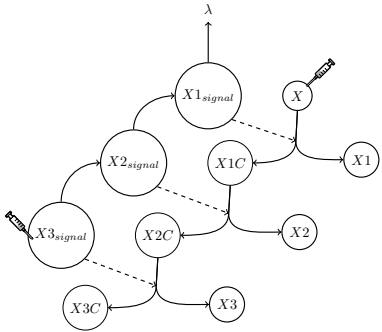}
	\caption{Time delay design with three stages. The syringe is used to indicate an injection of the input $X$ and the signal $X3_{signal}$. Species $X1$, $X2$, and $X3$ represent the output species from the delay line. Lambda ($\lambda$) shows decay of backwards propagation signal.}
	\label{fig:delayelement3}
\end{figure}

Extending the buffer for more than two stages is straightforward. Adding one output species ($Xm$), transition species ($XmC$), and catalyst species ($Xm_{signal}$) allows the system to flexibly provide a buffer of desired length. As an example, Figure~\ref{fig:delayelement3} shows how such a system would look having three delay stages in a back propagation delay line. The total number of species required in the system grows at a rate of $3m+1$, where $m$ is equal to the number of stages in the system. One trade-off to note is that as the number of stages in the system increases, so does the period of time to fully cascade the values of each species through the chain since each reaction ideally is idle and runs to full completion prior to $Xm_{signal}$ propagating backwards to begin the next conversion.

The reaction set of the delay line also scales in a straightforward fashion. Each step of delay has a reaction similar to reaction~\ref{eq:Xreac} for all intermediate phases with the final delay stage reaction (the $m^{th}$ delay) having a rate equation similar to reaction~\ref{eq:X1Creac}. This remains true for extending both the manual and backwards propagating delay line. Extension of the catalysts depends on the implementation. For the manual copy delay line, simply adding the species and a subsequent input is required. Extending the backward propagating delay line has the advantage that it does not increase the number of injections, but it still increases the overall number of species.

\section{Results}
\label{sec:results}
We will highlight the results for the two stage buffer and its extension beyond two stages. We employed Genetic Algorithms (GAs)~\cite{Melanie1999} to optimize the rate constants (mapped to chromosomes) of the backwards propagation model. We only used the algorithm to optimize the backwards propagation model since the manual copy was straightforward to optimize by hand. The GA used an elite selection of the top 20 chromosomes from the population of 100, which undergo cross-over and mutation to form the next generation. The goal (fitness function) of this evolutionary algorithm was to minimize the error of the delay line.

Error was defined as the difference between the actual input value ($X$) and the value occurring at $X1$ on this cycle and then $X2$ on the next cycle. This test was performed 50 time steps after $X$ is injected into the delay line. This equation shows the calculation of this error where $X[n]$ represents the current value of $X$ and $X[n-1]$ represents the value of $X$ on the previous input cycle.

\begin{equation}
error = |X_1 - X[n]| + |X_2 - X[n-1]|
\end{equation}

Adding the two errors for the two stage delay line provided the overall error. The genetic algorithm performed perturbation mutation that changed each chromosome's element with 30\% chance by $\pm$30\% using a uniform distribution. We ran the GA for 100 generations to produce the results for the two stage delay line. The algorithm was configured to target a transition of the input species, $X$, to the current time species, $X1$, as fast as possible, and convert the intermediate species, $X1C$ to the previous time species, $X2$, as fast as possible while minimizing the amount of leakage between the phases of the design.

\subsection{Two Stages}
The manual propagation delay line has the rate constants for each of the reactions as shown shown in Table~\ref{tab:rateconstantsMan}. Rates for the conversion of input species, $X$, down the chain is the same rate with the presence of $X1_{signal}$ and $X2_{signal}$ both increasing the rate by the same amount because the forward copy reactions should be as fast as possible. Figure~\ref{fig:delay2allMan} shows the plots using these rate constants in a two stage system, which can be replicated for a manual copy system of any size.

\begin{table}[h]
    \caption{Two stage manual copy GA rate constants.}
    \label{tab:rateconstantsMan}
    \centering
    \begin{tabular}{llll}
	    Reaction                                 & Forward Rate & $K_m$          \\ \hline
	    $X \xrightarrow{X1_{signal}} X1 + X1C$   & 0.0757       & 2.0000         \\
	    $X1C \xrightarrow{X2_{signal}} X2 + X2C$ & 0.0757       & 2.0000         \\
		$X2_{signal} \rightarrow \lambda$        & 0.5643       & (None)         \\
	    $X1_{signal} \rightarrow \lambda$        & 0.5643       & (None)         \\
    \end{tabular}
\end{table}

For a different size, the back propagation delay line has different rate constants. In addition, the rate constants were not grouped like the manual propagation delay line because it would drastically decrease the performance. Looking at the constants in Table~\ref{tab:rateconstants}, the reaction for species $X1C$ to $X2$ is the fastest. This is directly due to the rapid rate that $X2_{signal}$ is reacting to $X1_{signal}$. Effectively, to meet the first requirement of getting $X$ into $X1$ as fast as possible, the lower level transition of species (Reaction~\ref{eq:X1Creac}) must complete before. Figure~\ref{fig:delay2all} shows the output of a two stage with these results.

\begin{table}
    \caption{Two stage backwards propagation GA rate constants.}
    \label{tab:rateconstants}
    \centering
    \begin{tabular}{llll}
	    Reaction                                 & Forward Rate & $K_m$          \\ \hline
	    $X \xrightarrow{X1_{signal}} X1 + X1C$   & 0.0020       & 0.0225         \\
	    $X1C \xrightarrow{X2_{signal}} X2 + X2C$ & 0.0706       & 2.0000         \\
	    $X2_{signal} \rightarrow X1_{signal}$    & 1.3648       & (None)         \\
	    $X1_{signal} \rightarrow \lambda$        & 0.0039       & (None)         \\
    \end{tabular}
\end{table}

To compare the accumulated error of the two delay lines we used symmetric mean absolute percentage error (SAMP) is defined as
\begin{equation}
SAMP = 100 * \langle \frac{|y-\hat{y}|}{y+\hat{y}} \rangle,
\end{equation}
where $\langle.\rangle$ is the mean of a set of multiple values, $y$ is the actual value, and $\hat{y}$ is the expected value. We calculate an average SAMP per stage (unit size) by dividing cumulative SAMP with $m$. More specifically, using $n$ to represent a discrete time series sample and $m$ to represent the number of stages: 
\begin{equation}
SAMP = \frac{100}{m} * \sum \limits_{k=1}^{m} \langle \frac{|Xk-X[n-(k-1)]|}{Xk+X[n-(k-1)]} \rangle.
\end{equation}
Exemplified with two stages ($m=2$), gives this equation for SAMP
\begin{equation}
SAMP = \frac{100}{2} * \langle \frac{|X1-X[n]|}{X1+X[n]} + \frac{|X2-X[n-1]|}{X2+X[n-1]} \rangle \label{eq:sampn2}.
\end{equation}

We performed an evaluation of 10,000 tests each repeating the sequence of actions defined in Table~\ref{tab:as2Man} and Table~\ref{tab:as2} for 200 iterations (200,000 time steps). Using Equation~\ref{eq:sampn2} over 10,000 runs, Figure~\ref{fig:fiveStageAbsDiff} shows the results for a delay line of size two as well as for larger sizes (discussed in next section). The difference in values from ideal for the two stage delay line are quite small. This shows that for a two stage delay line, both types operate well. One thing to note is that the backwards delay line has a larger initial error which can accumulate over time.

\subsection{Over Two Stages}
\begin{figure}[!t]
	\centering
	\includegraphics[width=8cm]{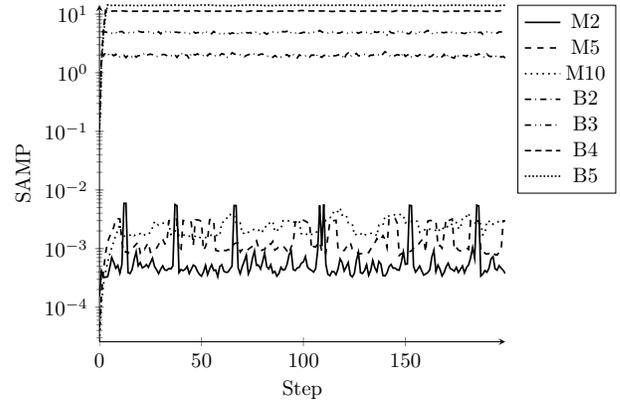}
	\caption{SAMP calculated for delay lines. M$m$ and B$m$ are the $m^{th}$ stage of manual copying and back propagation delay line.}
	\label{fig:fiveStageAbsDiff}
\end{figure}

In this section, we will examine the use of a delay line with five stages. Five stages was selected and executed for both the manual copy and back propagating delay line. Figure~\ref{fig:fiveStageAbsDiff} shows the final error when evaluated for 10,000 runs for 200 iterations each (same as for $m=2$ in previous section). The maximum error over the entire evaluation is shown in Table~\ref{tab:maxSAMP}. There are a few observations to note on this plot. The error on a backwards propagation delay line ($B$) increases as the number of stages in the delay line increases. For a smaller delay line, this error would generally be negligible, but for larger sizes this could be a concern. The manual copy has a significantly smaller error as shown in Figure~\ref{fig:fiveStageAbsDiff}.

\begin{table}[h]
	\centering
	\caption{Maximum and average SAMP obtained through performance runs of 200 iterations and varying configurations of stages and manual copy and backwards propagation. Maximum and average excludes the initial values where the delay line is filling (first $m$ points with low SAMP).}
	\label{tab:maxSAMP}
	\begin{tabular}{llllll}
	Backwards DL & Max     & Average & Manual DL & Max      & Average  \\ \hline
	B5           & 14.35\% & 14.09\% & M10       & 0.0059\% & 0.0016\% \\
	B4           & 11.66\% & 11.25\% & M5        & 0.0049\% & 0.0024\% \\
	B3           & 5.26\%  & 4.84\%  & M2        & 0.0033\% & 0.0008\% \\
	B2           & 2.28\%  & 1.97\%  & ~         & ~        & ~        \\
	\end{tabular}
\end{table}

As for the backwards propagating delay line, the error starts to accumulate to a noticeable value rapidly. Even by phase three, the delay line is starting to produce error that is in excess of the manual copy delay line with ten stages. Looking back to Figure~\ref{fig:delay2X2SignalZoomed} there is a period of time where both $X1_{signal}$ and $X2_{signal}$ overlap which can explain how error that starts quite small in stage one of the delay system accumulates to a large value by the time it reaches the later stages of the buffer. Depending on the desired properties of the delay line, this is worth considering for the application.

\subsection{Time Series Perceptron Integration}
\begin{figure}[h]
	\centering
	\includegraphics[width=5cm]{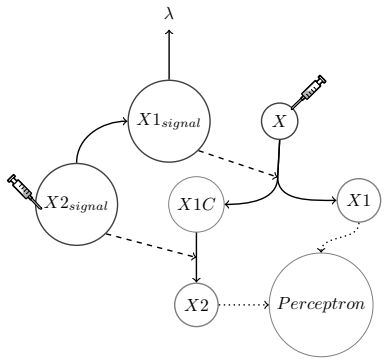}
	\caption{Perceptron integration with backwards propagating delay line of two stages. The delay line outputs ($X1$ and $X2$) are fed to the perceptron without modification of the delay line.}
	\label{fig:perceptintegrate}
\end{figure}

To demonstrate the capabilities of the delay line to fit into other designs, we integrated it with a chemical perceptron called a threshold asymmetric signal perceptron~\cite{Banda2014}. This perceptron learns through reinforcements and is inspired by biological neurons. Integration with the delay line and the perceptron shows how the delay line can easily fit with other systems to alter the type of input stream without any design modifications. Previously, the perceptron received both values simultaneously as two inputs. Now, we are showing that, without change to the perceptron or delay line, the two integrate together and function well. Figure~\ref{fig:perceptintegrate} shows an example of this integration.

\begin{figure}[h]
	\centering
	\includegraphics[width=8cm]{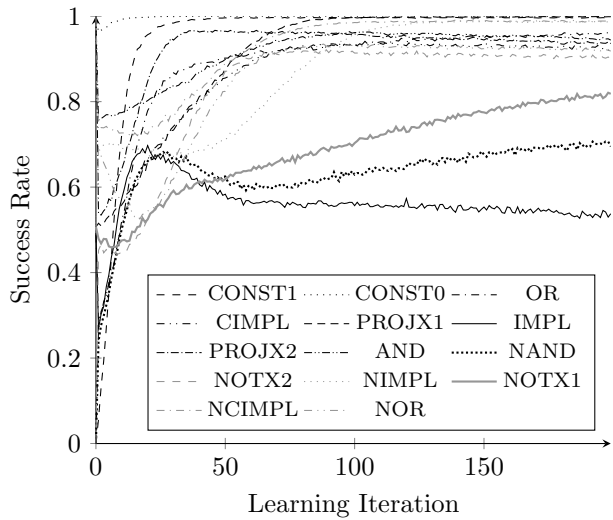}
	\caption{Binary time-series chemical perceptron success rate. The perceptron learns 11 of the 14 functions with an accuracy of greater than 85\%.}
	\label{fig:perceptLearning}
\end{figure}

\begin{figure*}[!t]
	\centering
	\begin{subfigure}[b]{0.45\textwidth}
		\centering
		\includegraphics[width=5cm]{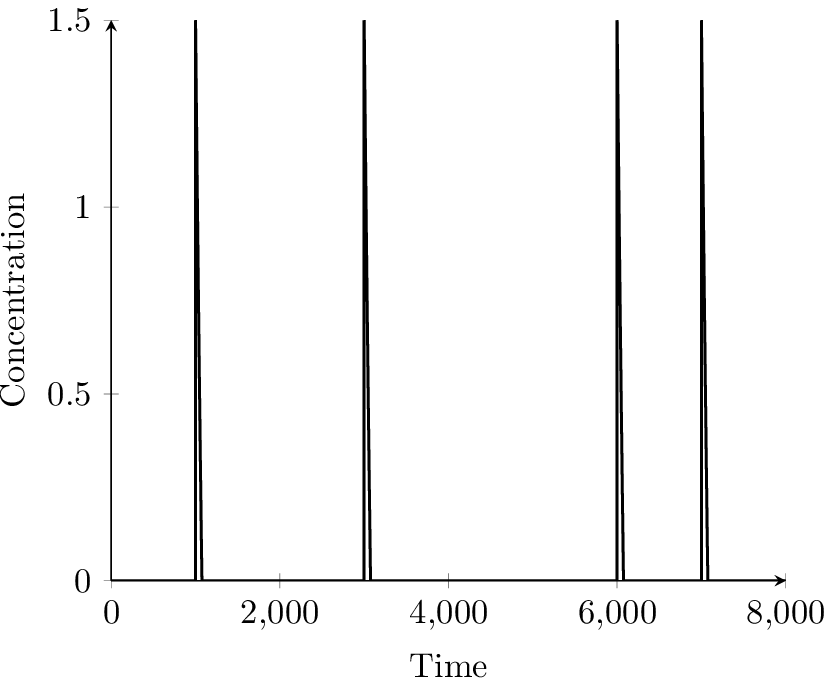}
		\caption{Input Stream}
		\label{fig:ORinput}
	\end{subfigure}
	\begin{subfigure}[b]{0.45\textwidth}
		\centering
		\includegraphics[width=5cm]{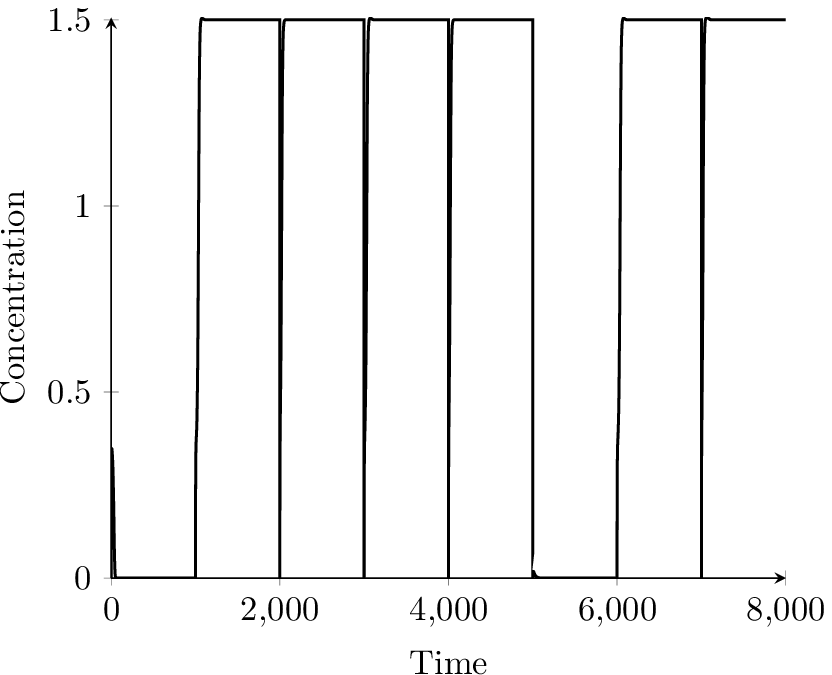}
		\caption{Output Stream}
		\label{fig:OROutput}
	\end{subfigure}
	\caption{Example concentration traces of binary time-series chemical perceptron that successfully learns OR function. Left shows input stream 0,1,0,1,0,0,1,1. Right shows correct output stream of 0,1,1,1,1,0,1,1. Two zeros on the input stream at 4,000 and 5,000 successfully produce zero at time 5,000 on output stream.}
	\label{fig:ORPerceptResults}
\end{figure*}

Using the reinforcement learning present in the perceptron~\cite{rojas1996neutral}, we trained it to attempt to learn 14 linearly separable binary functions. Figure~\ref{fig:perceptLearning} shows the results of this learning. The combined perceptron learns 11 of the 14 functions with an accuracy of greater than 85\%. Figure~\ref{fig:ORPerceptResults} shows the buffer and perceptron accurately producing the output for OR.

NAND, IMPL, and NOTX1 are all heavily dependent on the last input to resolve in the time delay line, $X1$. The input species $X1$ is not provided to the system until typically 50 time steps later than value $X2$. The original model of the perceptron was optimized for instantaneous and simultaneous injection of both inputs. Because input $X1$ is not ready, the performance is lower because that input plays a larger role on the correct performance for these logic functions. This makes the system capable of obtaining an average success rate of approximately 90\% compared to the perceptron's 99\% success rate~\cite{Banda2014}. The two combine well to form a binary time-series chemical perceptron.

\section{Deoxyribozyme Cascading Implementation}
\label{sec:deoxy}
\begin{figure*}[!t]
	\centering
	\includegraphics[width=12cm]{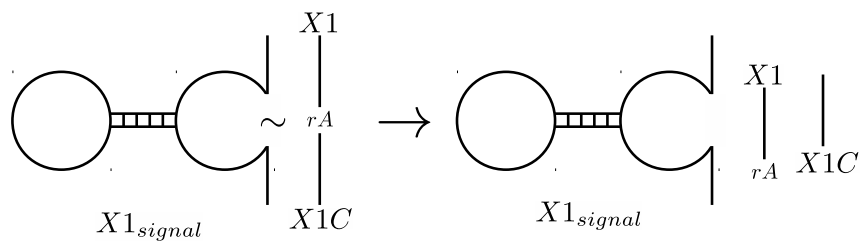}
	\caption{Deoxyribozyme cascading example. Deoxyribozyme $X1_{signal}$ cleaves $X$ at embedded ribonucleotide ($rA$) to form $X1$ and $X1C$. A similar process occurs on $X1C$ to produce $X2$ and $X2C$.}
	\label{fig:deoxy1}
\end{figure*}

Now, we would like to present how such a delay line could be realized in a system employing deoxyribozyme catalysis~\cite{stojanovic2003deoxyribozyme}~\cite{stojanovic2000homogeneous}~\cite{liu2009functional}. Figure~\ref{fig:deoxy1} shows an example of a two stage manual delay line with the signals being the deoxyribozymes $X1_{signal}$ and $X2_{signal}$ which cleave the substrate $X$ at the embedded ribonucleotide. This produces $X1$ ready for the next system to consume. Subsequently, $X1C$ embedded with another ribonucleotide is able to get cleaved by deoxyribozyme $X2_{signal}$ to form the next input to the system, $X2$.

\section{Conclusion}
\label{sec:conclude}
We have presented a novel implementation of a delay line as a chemical reaction network capable of storing past concentrations. Arranging our delay lines in a SIMD-like layout as shown earlier both show how this unit is able to delay multiple segments of data simultaneously with a shared control signal for either model of delay line. We have also demonstrated two different strategies for an implementation: manual copy delay line and backwards propagation delay line. A manual copy delay line can precisely carry values in a delayed state, but requires more intervention from the user (growing at a rate of $m$) to propagate values through the system. The second model, backwards propagating delay line, automatically moves values through the system with a single signaling injection with reasonable accuracy.

The integration of the backwards propagating delay line with the threshold asymmetric signal perceptron resulted in the first chemical model capable of learning binary time series. Also, this example is a proof-of-concept that our delay line is a modular block ready for use in other systems. For systems requiring a smaller window of past values, either model of the delay line gives sufficient accuracy for data storage. The manual copy delay line shows potential for longer chains with the amount of calculated SAMP passed between phases remaining below 0.01\% for a delay line of 10 stages. The backwards propagating delay line provides a much simpler user interface at the sacrifice of accuracy. A backwards propagation of five stages keeps the calculated SAMP below 15\%. Systems requiring a large number of delays will have to weigh accuracy and simplicity to make a selection for a particular implementation.

\section*{Acknowledgments}
This material is based upon work supported by the National Science Foundation under grant No. 1028120.

\bibliographystyle{IEEEtran}
\bibliography{bibliography}

\end{document}